\documentclass[aps,twocolumn,groupedaddress,floatfix]{revtex4-1}
\usepackage{amssymb,amsmath}

\usepackage{graphicx}
\usepackage{amsmath}
\usepackage{subfigure}
\usepackage[english]{babel}
\usepackage{float}
\usepackage{color}
\usepackage[document]{ragged2e}

\usepackage{lipsum}
\usepackage{suffix}
\usepackage{mathtools}
\DeclarePairedDelimiterX\MeijerM[3]{\lparen}{\rparen}%
{\begin{smallmatrix}#1 \\ #2\end{smallmatrix}\delimsize\vert\,#3}

\newcommand\MeijerG[8][]{%
  G^{\,#2,#3}_{#4,#5}\MeijerM[#1]{#6}{#7}{#8}}

\WithSuffix\newcommand\MeijerG*[7]{%
  G^{\,#1,#2}_{#3,#4}\MeijerM*{#5}{#6}{#7}}
\begin{document}
\newcommand{\be}{\begin{equation}}
\newcommand{\ee}{\end{equation}}
\newcommand{\rojo}[1]{\textcolor{red}{#1}}

\title{Fractionality and $\cal{PT}$- symmetry in a square lattice}

\author{Mario I. Molina}
\affiliation{Departamento de F\'{\i}sica, Facultad de Ciencias, Universidad de Chile, Casilla 653, Santiago, Chile}

\date{\today }

\begin{abstract} 
We study the spectral stability of a  2D discrete Schr\"{o}dinger equation on a square lattice, in the simultaneous presence of a fractional Laplacian and $\cal{PT}$ symmetry. For that purpose, we compute the plane-wave spectrum in closed form, as a function of the gain/loss parameter and the fractional exponent. Examination of the spectrum reveals that an increase of the gain/loss parameter favors the early appearance of complex eigenvalues, thus is, the onset of a broken ${\cal PT}$ symmetry. On the other hand, as the fractional exponent decreases from unity, at a critical value a gap opens up separating the upper and lower bands, and the spectrum becomes real. Further decrease of the exponent increases the width of the gap and the system remains in the $\cal{PT}$-symmetric phase down to a vanishing value of the fractional exponent. Examination of the density of states and the participation ratio reinforce these observations and lead one to conclude that, unlike the standard, non-fractional case where the binary lattice is always in the broken $\cal{PT}$ phase, for the fractional case it is possible to have a symmetric $\cal{P}{\cal T}$ phase in the presence of a finite gain/loss parameter and a small enough fractional exponent.

\end{abstract}

\maketitle

{\em Introduction}.  
Two physics developments have called for increased attention in recent times. One is the phenomenon of $\cal{P}{\cal T}$ symmetry, and the other is fractionality. Parity-time ($\cal{P}{\cal T}$) symmetric systems are characterized for having a non-hermitian Hamitonian, but a real spectrum nonetheless\cite{bender1,bender2}. This happens for a Hamiltonian that is invariant with respect to the simultaneous action of parity inversion and time reversal. In quantum mechanics the $\cal{P}{\cal T}$ symmetry conditions translate into the requirement that the real part of the potential be an even function in space $V_{R}(-x)=V_{R}(x)$, while the imaginary part be an odd function 
$V_{I}(-x)=-V_{I}(x)$. A particularly appropriate place for observing $\cal{P}{\cal T}$ symmetry is  optics, where the paraxial equation is akin to the Schr\"{o}dinger equation. There, the place of the `potential' is played by the complex index of refraction $n(x)=n_{R}+i\ n_{I}$, where $n_{R}(-x)=n_{R}(x)$ and $n_{I}(-x)=-n_{I}(x)$. Oftentimes the term $n_{I}$ is referred to as the `gain/loss parameter' since its presence can determine the energy gain or energy absorption of the system.
In the $\cal{P}{\cal T}$ regime a balanced gain and loss is possible.
In general, the spectrum of an optical $\cal{P}{\cal T}$ system remains real until the gain/loss parameter surpasses a critical value. At that point, two complex eigenvalues appear giving rise to  unstable dynamics. The ${\cal P}{\cal T}$ symmetry is then said to be spontaneously broken.

Currently, numerous $\cal{P}\cal{T}$-symmetric systems have been explored in several settings, from optics\cite{optics1,optics2,optics3,optics4,optics5}, electronic circuits\cite{circuits}, solid state and atomic physics\cite{solid1,solid2}, magnetic metamaterials\cite{MM}, among others. The ${\cal P}{\cal T}$ symmetry-breaking phenomenon has been observed in several experiments\cite{optics5,experiment2,experiment3}.

On the other hand, fractality is a subject that has evolved from a mathematical curiosity to a full-fledged area of research, Roughly speaking it is based on the idea of extending the normal integer derivative to one of non-integer order. Its beginning dates back to an old correspondence between Leibniz and L'Hopital where they examined some basic examples that seem to indicate that such extension could be done in principle, if one could fix potential consistency issues. A simple example is the derivative of a power function. For integer order we have $d^{n} x^{k}/d x^{n} = k!/(k-n)!\ x^{k-n}=\Gamma(k+1)/\Gamma(k-n+1)\ x^{k-n}$. Clearly, we can acommodate this to non-integer orders $d^{\alpha} x^{k}/d x^{\alpha} = \Gamma(k+1)/\Gamma(k-\alpha+1)\ x^{k-\alpha}$. Throughout the years, work by many people has produced formulas for non-integer derivatives that have proven useful in dealing with a variety of problems. One of the most used definitions is the Riemann-Liouville derivative
\be
\left( {d^{\alpha}\over{dx^{\alpha}}} \right) f(x) = {1\over{\Gamma(1-\alpha)}} {d\over{dx}} \int_{0}^{x} {f(s)\over{(x-s)^\alpha}}, 
\ee
where $0<\alpha<1$. The non-local character of the fractional derivative has proven useful in several fields:  Electrical propagation in cardiac tissue\cite{cardiac}, epidemics\cite{epidemics}, Levy processes in quantum mechanics\cite{levy}, fractional kinetics and anomalous diffusion\cite{kinetics1,kinetics2,kinetics3}, fluid mechanics\cite{quantum}, strange kinetics\cite{strange}, fractional quantum mechanics\cite{frac1,frac2}, plasmas\cite{plasmas}, biological invasions\cite{invasions}, among others.

While most of the studies involving fractionality are related to transport phenomena in continuous systems, with a continuous fractional Laplacian, interest on applications to discrete systems defined on a lattice have also aroused recent attention. This is so in part, to exact results obtained for 1D and 2D fractional discrete Laplacians\cite{roncal1,roncal2,molina1,molina2}. 

In this work we examine the mutual interplay between  ${\cal P}{\cal T}$ symmetry and fractality on a square lattice, focussing on the stability properties as a function of the gain/loss strength and the fractional exponent. In particular, we look at how the presence of nonlocal effects affects the energy gain and  loss balance of a system that obeys the ${\cal P}{\cal T}$ condition. As we can see, the presence of fractality serves to stabilize the system's spectrum. The density of states shows a strong tendency towards an increase in degeneration as the fractional exponent is decreased away from the standard case. The participation ratio of the modes also decreases signaling a degree of localization.
In general, we find that the presence of fractionality tends to restore the ${\cal P}{\cal T}$ symmetry to our binary lattice that, in the absence of fractality, is always in the broken ${\cal P}{\cal T}$ symmetry regime.

\begin{figure}[t]
 \includegraphics[scale=0.15]{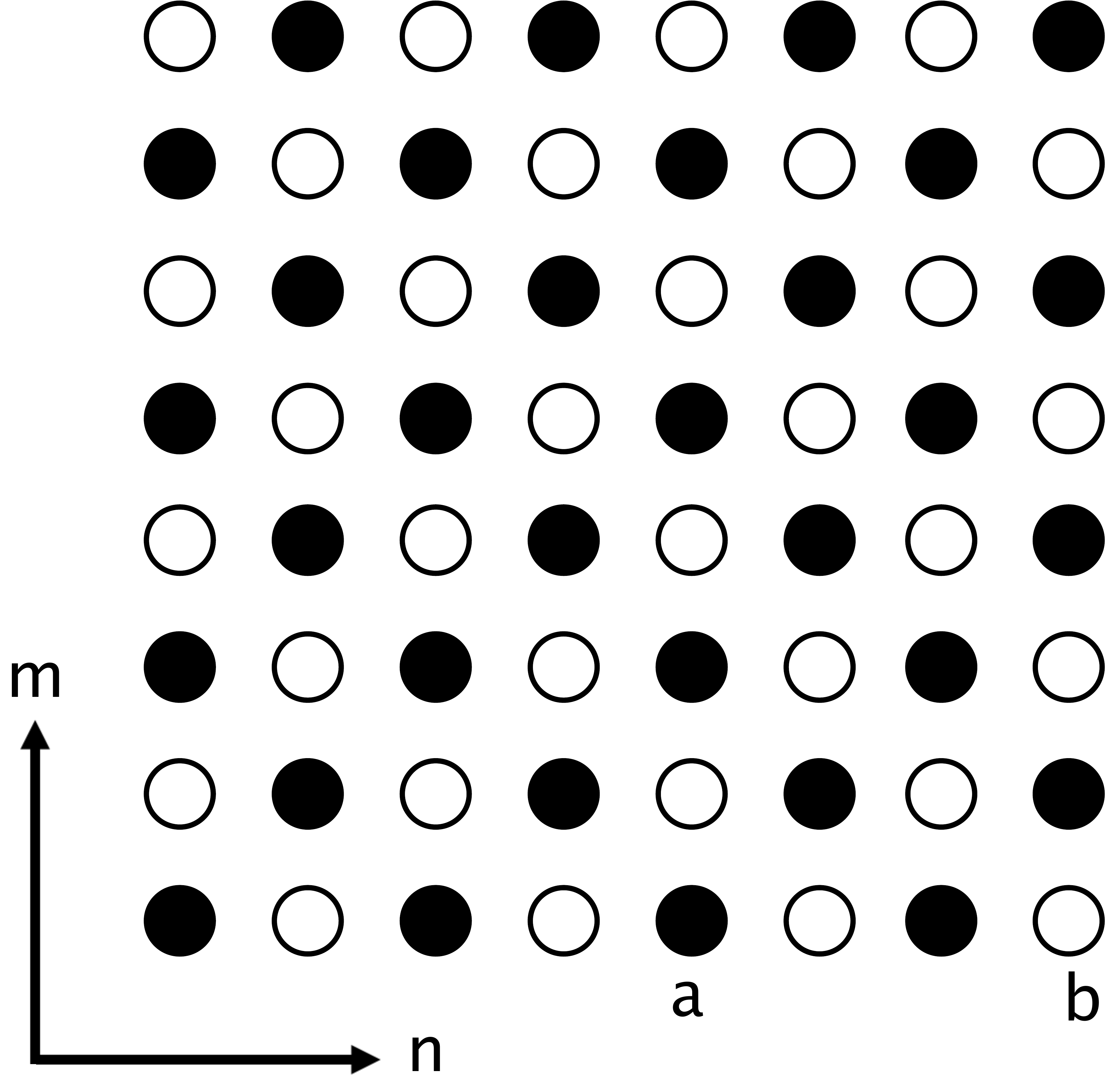}
  \caption{Square lattice with binary gain/loss distribution. 
  White (black) sites are endowed with gain (loss).}
  \label{fig1}
\end{figure}
{\em Model}.\ \ Let us consider a binary square lattice where at each site the imaginary part of the site energy can have the value $\gamma$ or $-\gamma$. The tight-binding equations are
\begin{eqnarray}
i {d c_{n m}\over{d t}} + i {(-1)^{n+m}} \gamma\ c_{n m} &+& V(c_{n+1,m}+c_{n-1,m}+\nonumber\\ & & c_{n,m+1}+c_{n,m-1})=0\label{eq2}
\end{eqnarray}
We decompose this into two interpenetrating lattices (Fig.1):
\begin{eqnarray}
i {d a_{n m}\over{d t}}+ i \gamma a_{n m} &+& V (b_{n,m+1}+b_{n,m-1}+\nonumber\\                                         & & b_{n+1,m}+b_{n-1,m})=0\label{2}
\end{eqnarray}
\begin{eqnarray}
i {d b_{n m}\over{d t}}- i \gamma b_{n m} &+& V (a_{n,m+1}+a_{n,m-1}+\nonumber\\                                          & & a_{n+1,m}+a_{n-1,m})=0\label{3}
\end{eqnarray}
As usual, we look for stationary state solutions: $a_{n m}(t)= a_{n m} e^{i \lambda t}$. Thus, we have
\begin{eqnarray}
(-\lambda+i \gamma) a_{n m} &+& V (b_{n,m+1}+b_{n,m-1}+\nonumber\\                                         & & b_{n+1,m}+b_{n-1,m})=0\label{4}
\end{eqnarray}
\begin{eqnarray}
(-\lambda-i \gamma) b_{n m} &+& V (a_{n,m+1}+a_{n,m-1}+\nonumber\\                                          & & a_{n+1,m}+a_{n-1,m})=0\label{5}
\end{eqnarray}
Before introducing fractality, let us look at the dispersion relation obtained from Eqs.(\ref{4}), (\ref{5}). We pose a plane-wave ansatz $a_{\bf n} = A\ e^{i {\bf k}\cdot {\bf n}}, 
b_{\bf n} = B\ e^{i {\bf k}\cdot {\bf n}}$, where ${\bf n}=(n_{1},n_{2})$ is a lattice site and ${\bf k}=(k_{1},k_{2})$ is a wavevector in the first Brillouin zone. We obtain:
\be
\lambda = \pm \left( -\gamma^2 + 4 V^2 (\ \cos(k_{1})+\cos(k_{2}))^2\  \right)^{1/2}.\label{6}
\ee
Inspection of Eq.(\ref{6}) reveals that, for any value of the gain/loss parameter $\gamma$, there are always ${\bf k}$ values for which $\lambda({\bf k})$ is imaginary. This is true even for arbitrarily small $\gamma$. The only exception is the trivial case $\gamma=0$. Thus, our bipartite square lattice  (\ref{4}),(\ref{5}) contains complex eigenvalues and is, therefore in the broken $\cal{PT}$ phase. 

As we will see next, the presence of fractality will have a stabilizing effect.

The non-fractional discrete Laplacian $\Delta_{n}$ is defined as 
\be
(\Delta_{\bf{n}}) f_{\bf{n}} = f_{n+1,m}+f_{n-1,m}-4 f_{n,m}+f_{n,m+1}+f_{n,m-1}
\ee 
In terms of $\Delta_{n}$, Eqs.(\ref{4}),(\ref{5}) can be rewritten as
\begin{eqnarray}
(-\lambda+i \gamma) a_{\bf n}+4 b_{\bf n}+ \Delta_{\bf n} b_{\bf n}&=&0\nonumber\\
(-\lambda-i \gamma) b_{\bf n}+4 a_{\bf n}+ \Delta_{\bf n} a_{\bf n}&=&0
\end{eqnarray}
We proceed now to replace the 2D discretized Laplacian by its fractional form $(\Delta_{n})\rightarrow (\Delta_{n})^s$, where $0<s<1$ is the order of the Laplacian and is known as the `fractional exponent'. When $s=1$ we recover the usual (non-fractional) discrete Laplacian. The explicit form for $(\Delta_{n})^s$ is given by\cite{roncal1,roncal2}
\begin{widetext}
\be
(\Delta_{n})^s f_{\bf j}= 
                        L_{2,s}\sum_{{\bf m}\neq {\bf j}} (f_{\bf m}-f_{\bf j})\  
\MeijerG*{2}{2}{3}{3}{1/2,-(j_2-m_2+1+s,j_2-m_2+1+s)}{1/2+s,j_1-m_1,-(j_1-m_1)}{1}\label{eq:9}
\ee
\end{widetext}
where 
${\bf j}=(j_1,j_2)$ and ${\bf m}=(m_1,m_2)$ are positions in the lattice, $G(...)$ is the  Meijer G-function and
\be
L_{2,s} = {4^s \Gamma(1+s)\over{\pi |\Gamma(-s)|}},
\ee
where $\Gamma(x)$ is the Gamma function.
An alternative expression for $(\Delta_{n})^s$ is
\be 
(\Delta_{n})^s f_{\bf n} = \sum_{{\bf m}\neq{\bf n} } (f_{\bf m} - f_{\bf n}) \ K^{s}({\bf n}-{\bf m})\label{eq:7}
\ee
where,
\be 
K^{s}({\bf m}) = {1\over{|\Gamma(-s)|}}\ \int_{0}^{\infty} e^{-4 t }\ I_{m_1}(2 t)\ I_{m_2}(2 t)\ t^{-1-s}\ dt \label{eq:8}
\ee
with ${\bf m} = (m_1,m_2)$ and $I_{m}(x)$ is the modified special Bessel function. Two limiting forms of $K^{s}({\bf m})$ will prove to be of importance: (a) $\lim_{s\rightarrow 1} K^{s}({\bf m}) = 1$ for ${\bf m}$ a nearest-neighbor of the origin, zero otherwise. (b) $\lim_{s\rightarrow 0} K^{s}({\bf m}) = {\cal O}(s)$.  

The stationary equations now read
\begin{eqnarray}
(-\lambda+ i \gamma) a_{\bf n}+4 V b_{\bf n}+V\ \sum_{\bf m} (b_{\bf m}-b_{\bf n}) K^s({\bf n}-{\bf m})&=&0\nonumber\\
(-\lambda- i \gamma) b_{\bf n}+4 V a_{\bf n}+V\ \sum_{\bf m} (a_{\bf m}-a_{\bf n}) K^s({\bf n}-{\bf m})&=&0.\ \ \ \ \ \ \ \ 
\end{eqnarray}
Now we pose the plane-wave ansatz: $a_{\bf n}= A\ e^{i {\bf k}\cdot{\bf n}}, b_{\bf n}= B\ e^{i {\bf k}\cdot{\bf n}}$. This leads to the system
\begin{gather}
(-\lambda+i \gamma)A + 4 V B +\nonumber\\
+ V B \sum_{\bf m}{(e^{i{\bf k}\cdot{({\bf m}-{\bf n}})}-1) K^s({\bf m}-{\bf n}})=0\nonumber\\
-(\lambda-i \gamma) B + 4 V A +\nonumber\\
V A \sum_{\bf m}{(e^{{\bf k}\cdot{({\bf m}-{\bf n}})}-1) K^s({\bf m}-{\bf n}})=0
\end{gather}

Imposing the vanishing of the determinant leads to the 
\begin{figure}[t]
 \includegraphics[scale=0.15]{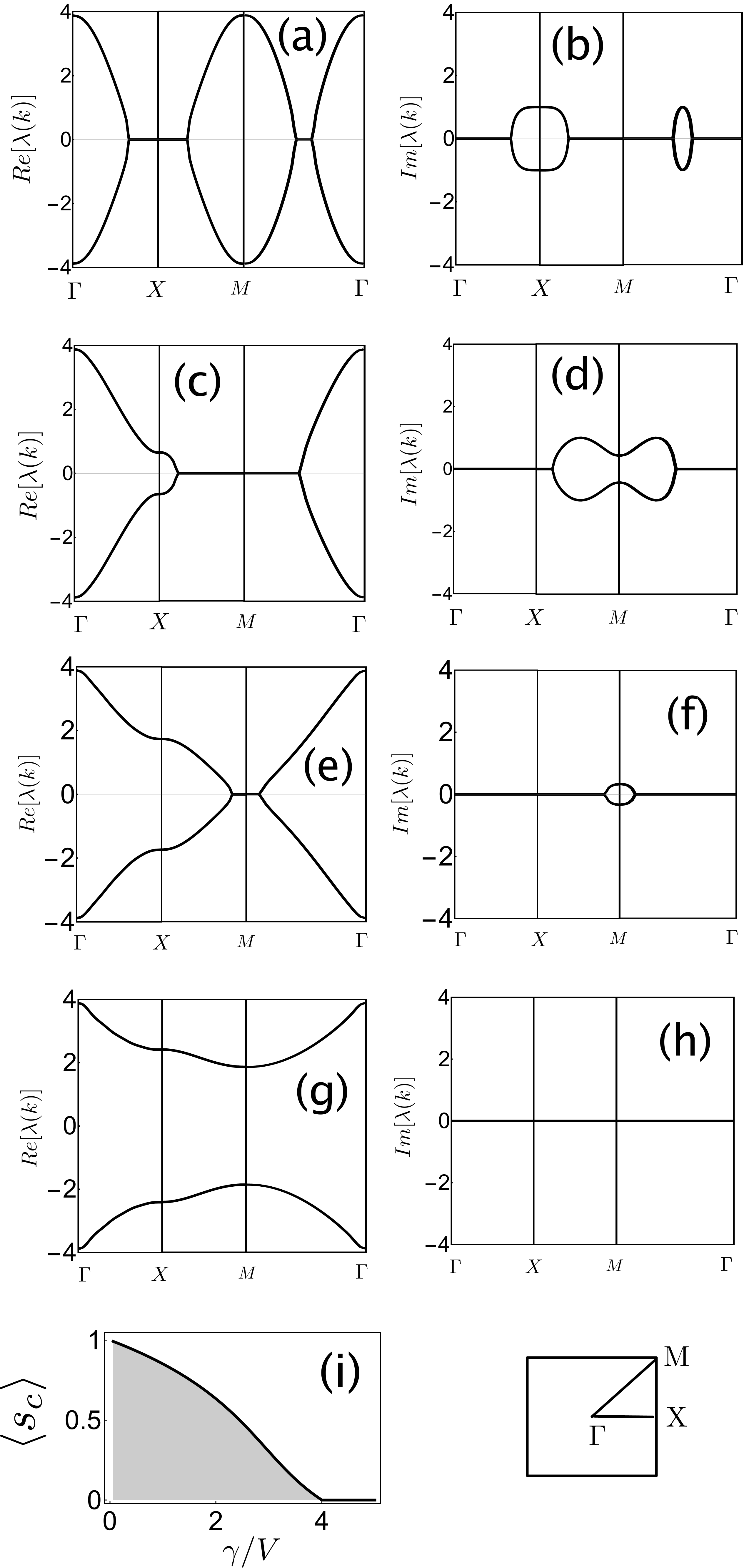}
  \caption{Real and imaginary parts of the dispersion relation (\ref{15}), plotted along specific directions inside the Brillouin zone, for different fractional exponents and a fixed gain/loss parameter value $\gamma=1$. (a) and (b): $s\approx 1$, (c) and (d): $s=0.8$, (e) and (f): $s=0.4$, (g) and (h): $s=0.2$ (i) Wavevector-averaged critical $s$ vs $\gamma$. Inside the shaded region the system remains in the symmetric ${\cal P T}$ phase.}
  \label{fig2}
\end{figure}
\begin{figure}[t]
 \includegraphics[scale=0.2]{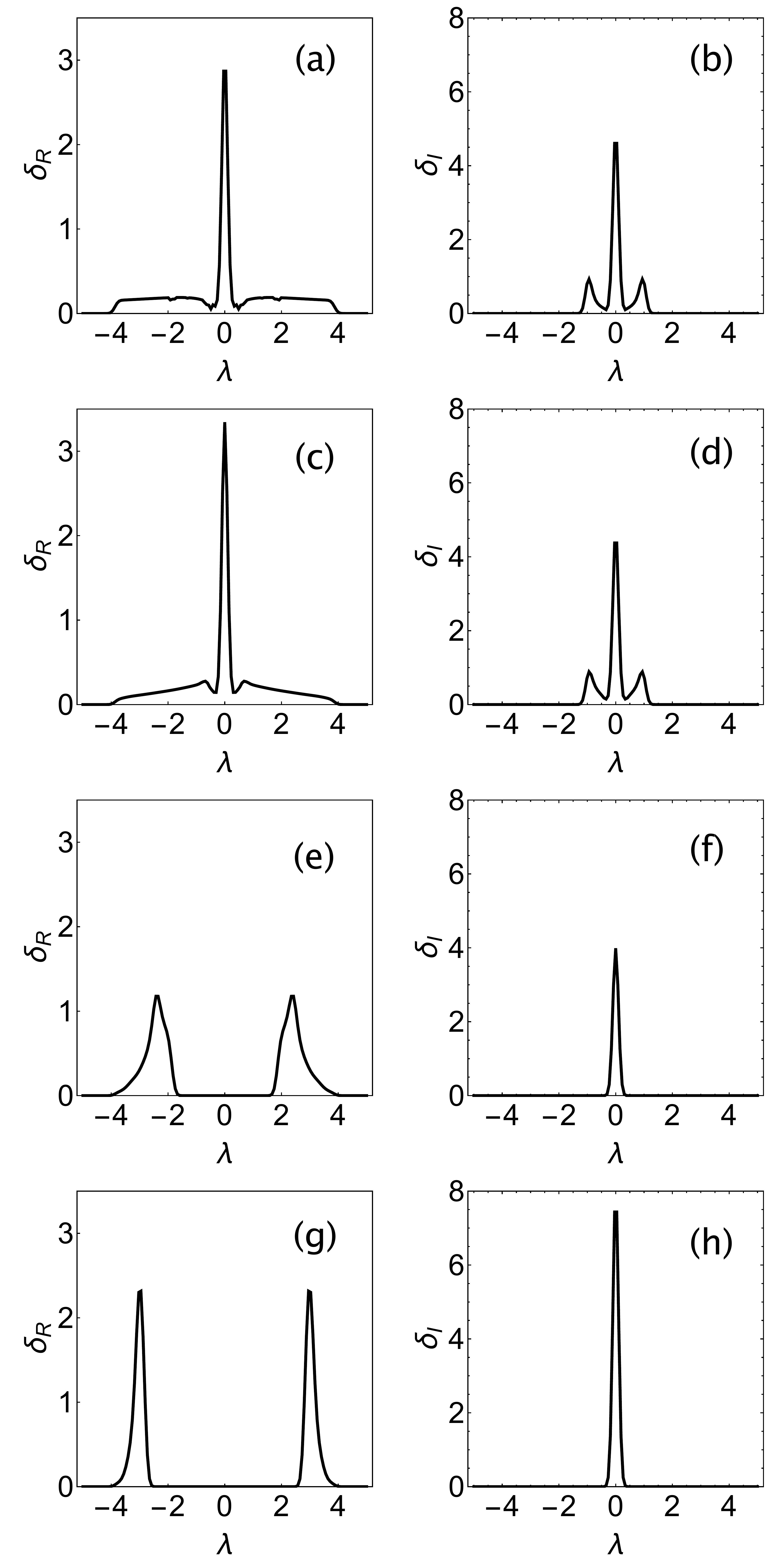}
  \caption{Left (right) column: Density of states of the real (imaginary) part of the spectrum, for $\gamma=1$. 
  (a) and (b): $s\approx 1$, (c) and (d): $s=0.8$, (e) and (f): $s=0.4$, (g) and (h): $s=0.2$}
  \label{fig3}
\end{figure}
dispersion relation
\be
\lambda({\bf k}) = \pm \sqrt{-\gamma^2 + V^2 \left( 4 + \sum_{\bf m}{(\cos{({\bf k}\cdot{\bf m}})-1)} K^s(\bf m) \right)^2 } \label{15}
\ee
For $s\rightarrow 1$, $K^s({\bf m})\rightarrow Z$, with $Z$ being the coordination number and one recovers the standard dispersion (\ref{6}) where the system is always in the broken ${\cal PT}$ phase. Now, we want to ascertain whether for $0<s<1$, it is possible to find a region(s) in parameter space $\{\gamma, s\}$ inside which $\lambda$ is purely real, that is, the system is in the symmetric ${\cal PT}$ phase leading to a bounded dynamics. From the general structure of $\lambda({\bf k})$, it is easy to see  that, for a fixed fractional exponent $s$,  an increase in the gain/loss parameter $\gamma$ favors the earlier appearance of complex eigenvalues, thus making the system unstable. In fact we observe that for $\gamma>4 V$, $\lambda(\bf k)$ is always imaginary regardless of $s, {\bf k}$. 
Let us focus on the more subtle effect of the fractional exponent. To that end, we proceed with a numerical sweep of $\lambda(\bf k)$ in $\{\gamma, s\}$ space. Representative results are shown in Fig.2 where we show the real and imaginary parts of $\lambda(\bf k)$ along standard directions inside the Brillouin zone, for several fractional exponents $s$ and a fixed gain/loss parameter $\gamma$. The very first plots (a) and (b) correspond to the non-fractional standard case $s=1$, and are included for reference.  From Eq.(\ref{15}) and $\gamma<4 V$ and as $s$ decreases away from unity, the two $\lambda({\bf k})$ branches separate at a given $s_{c}$ value where the imaginary part of the dispersion vanishes. For $0<s<s_{c}$, the band remains real and its band gap increases, reaching a value of $2 \sqrt{-\gamma^2+16  V^2}$ at $s\rightarrow 0$. As we can see, the imaginary branches first coalesce then shrinks and pop out of existence at $s=s_{c}$. On the other hand, the real branches seem to first coalesce and later split into a single pair at $s=s_{c}$. A quick, rough estimate of the critical value of the fractional exponent as a function of the gain/loss parameter $\gamma$ can be obtained from Eq.(\ref{15}), after averaging over the wavevector ${\bf k}$. This leads to the condition $\gamma \lesssim V(4-\sum_{\bf m}K^s({\bf m}))$, and is shown in Fig.2(i).

The density of states $\delta(\lambda)=(1/N) \sum_{\bf n} \delta(\lambda-\lambda(\bf k))$ is complex in general since $\lambda({\bf k})$ is either real or imaginary. We define partial densities of states for the real and imaginary part of the spectrum:
\begin{eqnarray}
\delta_{R}(\lambda) &=& (1/N) \sum_{\bf k} \delta(\lambda-Re[\lambda(\bf k)])\nonumber\\
\delta_{I}(\lambda) &=& (1/N) \sum_{\bf k} \delta(\lambda-Im[\lambda(\bf k)]).\label{16}
\end{eqnarray}
By using the analytical expression (\ref{15}) for $\lambda({\bf k})$ into Eq.(\ref{16}),
we compute numerically $\delta_{R}(\lambda)$ and $\delta_{I}(\lambda)$ for a fixed gain/loss value $\gamma=1$, and several fractional exponents $s$, ranging from $s\approx 1$ (standard case) down to $s\approx 0$. Results are shown in Fig.3. In agreement with Fig.2, we notice that as $s$ decreases, the imaginary part of the density of states decreases and vanishes altogether at some $s_{c}$ value, where it reduces to the sum of two delta-functions (rounded off in this case due to the finite lattice site). Density $\delta_{R}(\lambda)$ that keeps track of the density of states of the real part of the eigenvalues begins as a rather wide distribution at $s$ values much greater than zero; but as $s$ decreases and becomes smaller than $s_{c}$, it develops an energy split whose width increases with decreasing $s$, 
approaching $\delta(\lambda\pm 4 V)$ in the limit of vanishingly small $s$. For the density $\delta_{I}$ we have $\delta_{I}\rightarrow \delta(\lambda\pm i\,\gamma)$. 

The spatial extent of the modes is monitored through the participation ratio $PR$, defined as 
\be
PR = {\left(\sum_{n} |\phi_n|^2\right)^2\over{\sum_{n} |\phi_n|^4}}
\ee
where $\phi_{n}$ is the amplitude at site $n$ of a given stationary mode $\phi$. We find the eigenmodes by solving the eigenvalue problem stemming from Eq.(\ref{eq2}) for a finite $N\times N$ lattice. For a completely extended mode, $PR=N\times N$ and for a completely localized one, $PR=1$. It is interesting to see how the interplay between $\gamma$ and $s$ 
\begin{figure}[t]
 \includegraphics[scale=0.26]{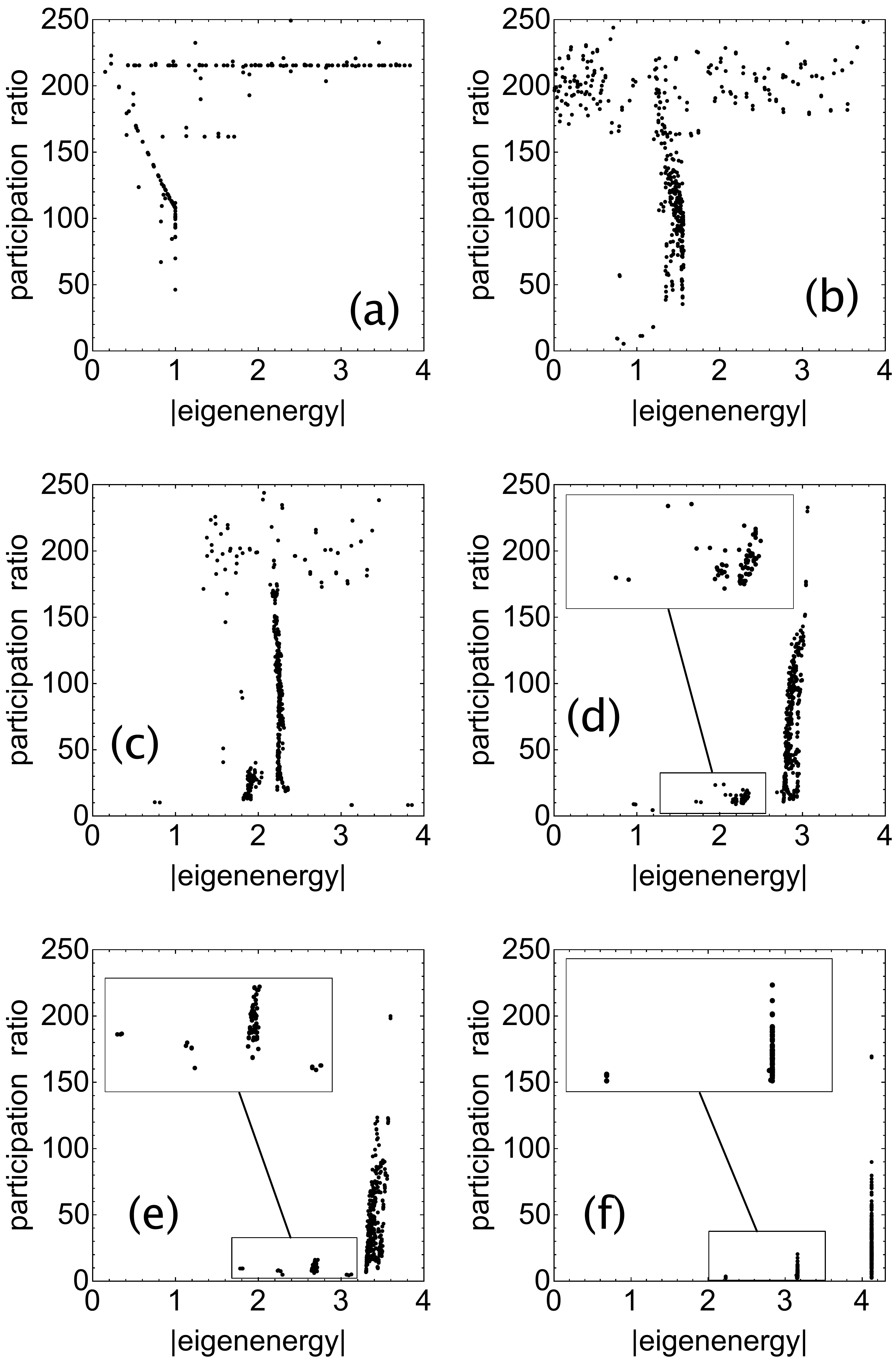}
  \caption{Participation ratio of all modes as a function of the magnitude of their eigenenergies for  $N\times N=441$ sites, $\gamma=1$ and several different $s$ values: (a) $s\approx 1$, (b) $s=0.8$, (c) $s=0.6$, (d) $s=0.4$, (e) $s=0.2$ and (f) $s\approx 0$. The insets show augmented vistas of some regions close to the horizontal axis.}
  \label{fig4}
\end{figure}
can affect $PR$. Thus, for a given pair $(\gamma,s)$ we compute the PR of all modes and do a scatter plot as a function of the absolute value of the mode eigenenergy. We will consider here a finite $N\times N$ square lattice in order to underlie certain features of the PR in the limit of vanishing fractional exponent. Figure 4 shows $PR$ vs the absolute value of the mode eigenvalue, for a fixed $\gamma=1$ and several fractional exponents $s$ ranging from $s\approx 1$ which is the standard, non-fractional case, down to $s\approx 0$. For most fractional exponents, we notice a tendency for the mode energies to cluster around certain energies and for the PR to span a range of values, going from a completely localized mode ${\cal O}(1)$ up to a more delocalized one ${\cal O}(N)$. In the limit $s\rightarrow 0$, $K^s({\bf m})\rightarrow 0$ and the stationary equations reduce to $(-\lambda + i \gamma_{\bf n}+ Z V) C_{\bf n}\approx 0$, where $Z$ is the coordination number: $Z=2$ at the corner, $Z=3$ at the edge and $Z=4$ at the bulk of the square lattice. This leads to the possible eigenvalues:
$\lambda=2 V \pm i \gamma$ which appears $4$ times, $\lambda=3 V \pm i \gamma$ which appears $4 (N-2)$ times, and $\lambda=4 V \pm i \gamma$ which appears $(N-2)^2$.  Note that for an infinite square lattice, the bulk value $4 V$ will dominate.  Thus, for $s=0$ the absolute values of the eigenvalues converge to $|Z \pm i \gamma|$. As an example, for the $N\times N = 441$ case shown in Fig.4 the absolute value of the degenerate energies for $\gamma=1$ are
$\sqrt{4^2+1}=4.12$ (with degeneracy $361$ ), $\sqrt{3^2+1}=3.16$ (with degeneracy $76$) and $\sqrt{2^2+1}=2.24$ (with degeneracy $4$), which are clearly shown in Fig. 4(f). 

{\em Discussion.}\ As mentioned in the Introduction, the usual 2D bipartite square lattice is always in the broken ${\cal PT}$ phase for any gain/loss parameter. The symmetry can, however, be restored if a strain in the couplings is applied\cite{strain}. In this paper we follow a different route by introducing fractionality into the model. We are thus faced with the interplay between two ingredients: ${\cal PT}$ and fractionality. The dispersion relation of the model was computed in closed form, showing a regime in fractional exponent where the spectrum is completely real, provided the gain/loss parameter is smaller than threshold. The density of states shows the onset of a gap between the two,  real energy branches as well as the onset of degeneracy at low values of fractionality. The participation ratio, computed for a finite lattice, shows that for small $s$ values, the energy of the modes tend to cluster around three energy values, each one with a different degeneracy. The values of these degeneracies depend on boundary effects, while the existence of localized modes can be traced back to the formation of a flat band in the limit $s\rightarrow 0$ for a large system.

In  summary, we might say that the main effect of an increase of the gain/loss parameter, is to bring the system closer to the broken ${\cal{PT}}$ symmetric regime, while a decrease of the fractional exponent   tends to restore the ${{\cal PT}}$-symmetry and increase the mode degeneracy which becomes complete at $s\rightarrow 0$. The existence of a stabilizing mechanism is quite interesting and potentially useful in general since it allows to reach a regime of high gain/loss that is otherwise unstable.  This stabilizing effect on the ${\cal PT}$ symmetric phase could be observed, for instance,  in discrete optical systems like coupled waveguide arrays, where fractionality can be emulated\cite{emulate}.

\acknowledgments
This work was supported by Fondecyt Grant 1200120.

\end{document}